\documentclass[12pt]{article}
\usepackage{amsmath,latexsym,amssymb,epsfig,psfrag}  
\usepackage{comment}

\def\ltsim{\lower3pt\hbox{$\, \buildrel < \over \sim \, $}}  
\def\gtsim{\lower3pt\hbox{$\, \buildrel > \over \sim \, $}}  
\newcommand{\be}{\begin{equation}}  
\newcommand{\ee}{\end{equation}}  

\newcommand{\g}{\gamma}
\newcommand{\bea}{\begin{eqnarray}}
\newcommand{\eea}{\end{eqnarray}}
\def\ga{\mathrel{\raise.3ex\hbox{$>$\kern-.75em\lower1ex\hbox{$\sim$}}}}  
\def\la{\mathrel{\raise.3ex\hbox{$<$\kern-.75em\lower1ex\hbox{$\sim$}}}}

\makeatletter   
\@addtoreset{equation}{section}   
\makeatother

\begin{document}  
  
\baselineskip=16pt   
\begin{titlepage}  
\begin{center}  
  
\vspace{0.5cm}  
  
\large {\bf LARGE-SCALE CURVATURE PERTURBATIONS}
\vskip 0.1cm
\large {\bf WITH SPATIAL AND TIME VARIATIONS}
\vskip 0.1cm
\large {\bf OF THE INFLATON DECAY
RATE}

\vspace*{10mm}  
\normalsize  

{\bf S. Matarrese\footnote{sabino.matarrese@pd.infn.it} and 
A.~Riotto\footnote{antonio.riotto@pd.infn.it}}   

\smallskip   
\medskip   
\it{Department of Physics and INFN,\\
Sezione di Padova, via Marzolo 8,
I-35131 Padova, Italy}


\vskip0.6in \end{center}  
   
\centerline{\large\bf Abstract}
\vskip 1cm  
\noindent
We present a gauge-invariant formalism to 
study the evolution of the 
curvature and entropy perturbations in the case in which spatial
and time variations of the inflaton decay rate into ordinary
matter are present. During the reheating stage after inflation  
curvature perturbations can vary with time on super-horizon scales 
sourced by a 
a gauge-invariant inflaton decay rate perturbation. We show
that the latter is a function not only of the spatial variations
of the decay rate generated during inflation, as envisaged
in a recently proposed scenario, but also of
the time variation of the inflaton decay rate during 
reheating.  If only the second source is present, the final curvature
perturbation at the end of the reheating stage is proportional
to the curvature perturbation at the beginning of reheating with 
a coefficient of proportionality which can be either smaller
or larger than unity depending upon the underlying
physics governing the time variation of the inflaton decay rate.
As a consequence, we show that 
the standard consistency relation between the amplitude of 
curvature perturbations, the amplitude of tensor perturbations and the 
tensor spectral index of one-single field models of inflation is violated
and there is 
the possibility that the tensor-to-curvature amplitude 
ratio is larger than in the standard case.

\vspace*{2mm}   

\vskip 1cm
\begin{flushleft}
DFPD-TH/03/24
\end{flushleft} 
\end{titlepage}  
  
\section{Introduction}  \label{sec:intro}


One of the basic ideas of modern cosmology is that there was an epoch early
in the history of the universe when potential, or vacuum, energy 
associated to a scalar field, the inflaton $\phi$, 
dominated other forms of energy density such as matter or radiation. 
During such a
vacuum-dominated era the scale factor grew exponentially (or nearly
exponentially) in time. During this phase, dubbed inflation 
\cite{guth81,lrreview},
a small,  smooth spatial region of size less than the Hubble radius
could grow so large as to easily encompass the comoving volume of the 
entire presently observable universe. If the universe underwent
such a period of rapid expansion, one can understand why the observed
universe is so homogeneous and isotropic to high accuracy.

Inflation has also become the dominant 
paradigm for understanding the 
initial conditions for structure formation and for Cosmic
Microwave Background (CMB) anisotropy. In the
inflationary picture, primordial density and gravity-wave fluctuations are
created from quantum fluctuations ``redshifted'' out of the horizon during an
early period of superluminal expansion of the universe, where they
are ``frozen'' \cite{muk81,hawking82,starobinsky82,guth82,bardeen83}. 
Perturbations at the surface of last scattering are observable as temperature 
anisotropy in the CMB which was first detected by the Cosmic Background 
Explorer (COBE) satellite \cite{smoot92,bennett96,gorski96}.
The last and most impressive confirmation of the inflationary paradigm has 
been recently provided by the data 
of the Wilkinson Microwave Anistropy Probe (WMAP) mission which has 
marked the beginning of the precision era of the CMB measurements in space
\cite{wmap1}.
The WMAP collaboration has  produced a full-sky map of the angular variations 
of the CMB, with unprecedented accuracy.
WMAP data confirm the inflationary mechanism as responsible for the
generation of curvature (adiabatic) super-horizon fluctuations. 

Despite the simplicity of the inflationary paradigm, the mechanism
by which  cosmological adiabatic perturbations are generated  is not
yet fully established. In the standard picture, the observed density 
perturbations are due to fluctuations of the inflaton field itself. 
When inflation ends, the inflaton oscillates about the minimum of its
potential and decays, thereby reheating the universe. As a result of the 
fluctuations
each region of the universe goes through the same history but at slightly
different times. The 
final temperature anisotropies are caused by the fact that
inflation lasts different amounts of time in different regions of the universe
leading to adiabatic perturbations. Under this hypothesis, 
the WMAP dataset already allows
to extract the parameters relevant 
for distinguishing among single-field inflation models \cite{ex}.

An alternative to the standard scenario is represented by the curvaton 
mechanism
\cite{curvaton1,curvaton2,curvaton3} where the final curvature perturbations
are produced from an initial isocurvature perturbation associated to the
quantum fluctuations of a light scalar field (other than the inflaton), 
the curvaton, whose energy density is negligible during inflation. The 
curvaton isocurvature perturbations are transformed into adiabatic
ones when the curvaton decays into radiation much after the end 
of inflation. 

Recently, another mechanism for the generation of cosmological
perturbations has been proposed \cite{gamma1,gamma2,gamma3}.  
It acts during the reheating
stage after inflation and it was dubbed the ``inhomogeneous reheating'' 
mechanism in Ref. \cite{gamma3}. The coupling of the inflaton to normal matter 
may be  determined by the vacuum expectation value of fields $\chi$'s
of the underlying theory. 
If those fields are light during inflation, 
fluctuations  $\delta\chi\sim\frac{H}{2\pi}$,
where $H$ is th Hubble rate during inflation, are left imprinted
on super-horizon scales. These perturbations
lead to spatial 
fluctuations in the decay rate $\Gamma$ 
of the inflaton field to ordinary 
matter, $\frac{\delta\Gamma}{\Gamma}
\sim\frac{\delta\chi}{\chi}$,
causing  adiabatic perturbations
in  the final reheating temperature
in different regions of the universe. 

Interestingly, these different  scenarios 
have different observational predictions. 
The curvaton scenario and the one based on variation of the
decay rate 
allows to generate the observed level of density perturbations with a 
much lower scale of inflation and thus generically predicts 
a smaller level of gravitational waves. 
Furthermore, because the field responsible for the fluctuations is not 
the inflaton, 
it can have significantly larger self couplings and thus density perturbations
could be non-Gaussian. The non-Gaussianity can be large enough to be 
detectable by CMB and Large Scale Structure observations contrary
to what predicted in the traditional one-single field model
of inflation, where the level of non-Gaussianity is very small \cite{ng}.

Contrary to the standard picture, both the curvaton and the 
inhomogeneous reheating mechanism exploit the fact that 
the total curvature perturbation (on uniform density hypersurfaces)
$\zeta$ can change on arbitrarily large scales due to a non-adiabatic
pressure perturbation    which may be 
present  in a multi-fluid system \cite{Mollerach,MFB,GBW,WMLL}.
While the entropy
perturbations evolve independently of the curvature perturbation on
large scales,  the evolution of the large-scale curvature is
sourced by entropy perturbations.

In this paper, we present a gauge-invariant formalism to 
study the evolution of the 
gauge-invariant curvature and entropy perturbations in the 
inhomogeneous reheating mechanism. 
To do so, we extend the gauge-invariant formalism first introduced in 
Ref. \cite{wmu}, which is appropriate to describe the evolution of curvature 
and entropy perturbations in multi-fluid cosmologies when energy transfer 
between fluids is included. We show
that the curvature perturbation during the reheating stage
can vary with time on super-horizon scales sourced by a 
a gauge-invariant inflaton decay rate perturbation

\begin{equation}
\delta\Gamma^{\rm GI}=\delta\Gamma-\dot\Gamma
\frac{\delta\rho_\phi}{\dot\rho_{\phi}}\,,
\end{equation}
where $\rho_\phi$ is the inflaton energy density and $\delta\rho_\phi$
its perturbation.
The gauge-invariant inflaton decay rate perturbation
is a function not only of  the spatial variation
of the decay rate $\delta\Gamma$ generated during inflation, but also of
the time variation of the inflaton decay rate $\dot\Gamma$. To our knowledge,
this new  source proportional to $\dot\Gamma$ has never been discussed
in the literature.

As an application, we study the evolution of the curvature perturbation
in the  inhomogeneous reheating scenario
where fluctuations of the inflaton decay rate $\delta\Gamma$ are induced by 
the fluctuations of some light scalar field during inflation and
confirm the results of Refs. \cite{gamma1,gamma3}.  We extend their findings
showing that variation of the total curvature perturbation $\zeta$
on super-horizon scales  take place even when the inflaton decay rate
changes with time during reheating. This new effect is
proportional to the total curvature perturbation
generated during the inflationary 
period $\zeta_{\rm in}$  and causes either an increase 
or a depletion of $\zeta_{\rm in}$ depending upon the underlying
model responsible for the time variation of $\Gamma$. Furthermore,
we show that the standard consistency relation between the amplitude of 
curvature perturbations, the amplitude of tensor perturbations and the 
tensor spectral index is violated. 
Finally, for completeness, we
extend our analysis to the case in which
the classical inflaton field $\phi$ does not release its energy 
perturbatively, but 
very rapidly (explosively) decays into either its own quanta  
or into other bosons due to broad parametric resonance, the so-called
preheating stage.

\section{The perturbative inflaton decay} 
Our starting point are the   equations governing the
evolution of cosmological perturbations during the reheating stage.
We follow the gauge-invariant approach developed in Ref. \cite{wmu}
for the general
case of an arbitrary number of interacting fluids in general
relativity.

At the end of inflation,  once the Hubble rate drops below
the mass of the inflaton field $\phi$, the inflaton 
starts oscillating around the minimum of its potential.
Averaged over several oscillations, the effective equation of state is 
$\langle P_\phi/\rho_\phi \rangle=0$, where $P_\phi$ and $\rho_\phi$ 
are the inflaton pressure and energy density, respectively.  
The coherent oscillations of the inflaton field are equivalent to a 
fluid of non-relativistic particles \cite{Turner}. 
The vacuum energy during inflation is transformed into the energy density of 
the coherent inflaton oscillations.
Assuming the inflaton is unstable and decays into light particles
(``radiation'') with a decay rate $\Gamma$, this represents an energy
transfer from the pressureless inflaton fluid to the radiation fluid
with energy density $\rho_\gamma$. 

The evolution of the background FRW universe during the reheating stage
is governed by the
Friedmann constraint
\begin{eqnarray}
\label{Friedmann}
H^2 &=& \frac{8\pi G}{3}\rho \,,
\\
\dot H &=& -4\pi G \left( \rho+P\right)
 \,,
\end{eqnarray}
and the continuity equation
\begin{equation}
\label{continuity}
\dot\rho=-3H\left( \rho+P\right)\,,
\end{equation}
where the dot denotes differentiation with respect to the coordinate time
$t$, $H\equiv \dot a/a$ is the Hubble parameter, and
$\rho$ and $P$ are the total energy density and the total
pressure of the system.
The total energy density
and the total pressure are related to the energy density and
pressure of the inflaton field and radiation  by

\begin{eqnarray}
\rho &=&\rho_\phi+\rho_\gamma \,, \nonumber\\
P&=& P_\phi+P_\gamma \,, 
\end{eqnarray}
where $P_\gamma$ is the radiation pressure. 
The inflaton field $\phi$ and the radiation component 
have energy-momentum tensor $T^{\mu\nu}_{(\phi)}$ and 
$T^{\mu\nu}_{(\gamma)}$, respectively.
The total energy momentum tensor $T^{\mu\nu}=T^{\mu\nu}_{(\phi)}+
T^{\mu\nu}_{(\gamma)}$ is covariantly conserved, but we allow for
energy transfer between the fluids,
\begin{eqnarray}
 \label{Qvector}
\nabla_\mu T^{\mu\nu}_{(\phi)}&=&Q^\nu_{(\phi)}\,,\nonumber\\
\nabla_\mu T^{\mu\nu}_{(\gamma)}&=&Q^\nu_{(\gamma)}\,, 
\end{eqnarray}
where $Q^\nu_{(\phi)}$ and $Q^\nu_{(\gamma)}$ are
 the generic energy-momentum transfer to
the inflaton and radiation sector respectively
and are  subject to the constraint
\begin{equation}
\label{Qconstraint}
Q^\nu_{(\phi)}+Q^\nu_{(\gamma)}=0 \,.
\end{equation}
The continuity equations for
the energy density of the inflaton field $\rho_\phi$ and radiation 
$\rho_\gamma$ in the background is thus \cite{KS} ($Q_\phi=Q_{(\phi)}^0$, 
$Q_\gamma=Q_{(\gamma)}^0$)
\begin{eqnarray}
\dot\rho_{\phi}
&=&-3H\left(\rho_{\phi}+P_{\phi}\right) +Q_{\phi}\,,\nonumber\\
\dot\rho_{\gamma}
&=&-3H\left(\rho_{\gamma}+P_{\gamma}\right) +Q_{\gamma}\,.
\label{m}
\end{eqnarray}

From now on, we parametrize, the 
energy transfer from the inflaton to radiation
by \cite{wmu}
\begin{eqnarray}
\label{defbackQa}
Q_{\phi} &=& -\Gamma\rho_{\phi} \,, \nonumber\\
\label{defbackQb}
Q_{\gamma} &=& \Gamma\rho_{\phi} \,,
\end{eqnarray}
where $\Gamma$ is the decay rate of the inflaton into
radiation. The positions (\ref{defbackQb}) are valid in the case in which
the inflaton decays into light states through a perturbative
process. However, if the classical inflaton field $\phi$ 
very rapidly (explosively) decays into either its own quanta  
or into other bosons due to broad parametric resonance, the so-called
preheating stage \cite{pre}, Eqs. (\ref{defbackQb}) should be modified.
The corresponding equations will be discussed in Section
\ref{sec:pre}.

The energy conservation equations are therefore
\begin{eqnarray}
\label{dotrhos}
\dot\rho_\phi &=& -\rho_\phi\left(3H+\Gamma\right)\,, \\
\label{dotrhog}
\dot\rho_\gamma &=& -4H\rho_\gamma+\Gamma\rho_\phi\,, \\
\label{dotrhotot}
\dot\rho&=&-H\left(3\rho_\phi+4\rho_\gamma\right) \,.
 \end{eqnarray}
It is
convenient to work in terms of the dimensionless density
parameters \cite{wmu}
\begin{equation}
\Omega_\phi \equiv \frac{\rho_\phi}{\rho}\,,\quad
\Omega_\gamma \equiv \frac{\rho_{\gamma}}{\rho}\,,
\end{equation}
and the dimensionless ``reduced'' decay rate \cite{wmu}
\begin{equation}
\quad g\equiv\frac{\Gamma}{\Gamma+H} \,,
\end{equation}
which varies monotonically from $0$ to $1$ in an expanding
universe.

The background equations (\ref{dotrhos}-\ref{dotrhotot}) can
then be written as an autonomous system
\begin{eqnarray}
\label{omegasprime}
 \Omega_\phi' &=& \Omega_\phi
 \left(\Omega_\gamma-{g\over1-g}\right) \label{din3} \,,\\
 \Omega_\gamma' &=& \Omega_\phi
 \left({g\over1-g}-\Omega_\gamma\right) \label{din1}\,,\\
\label{gprime}
 g' &=& {1\over2}(4-\Omega_\phi)(1-g)g +\frac{\Gamma^\prime}{\Gamma}
g(1-g)\label{din4}\,,
\end{eqnarray}
where the prime denotes differentiation with respect to the
number of e-foldings $N\equiv \ln a$, and we have allowed for a
time variation of the inflaton decay rate.
The density parameters are subject to the constraint
\begin{equation}
 \Omega_\phi + \Omega_\gamma = 1 \label{con1}\,.
\end{equation}

\subsection{Gauge-invariant perturbations}\label{sec:gauge}

Linear scalar perturbations about a
spatially-flat FRW background model are defined by the line
element

\begin{equation} 
ds^2=-(1+2\varphi)dt^2+2aB_{,i}dt dx^i
+a^2\left[(1-2\psi)\delta_{ij}+2E_{,ij}\right]dx^idx^j \,, 
\end{equation}
where we have used the notation of Ref. \cite{MFB} for the
gauge-dependent curvature perturbation, $\psi$, the lapse
function, $\varphi$, and scalar shear, $\chi\equiv a^2\dot E - aB$.

The zero-th component of the   
perturbed energy transfer vectors, Eq.~(\ref{Qvector}),
including terms up to first order, are  written as~\cite{KS}
\bea
 Q_{(\phi)0} &=& -Q_{\phi}(1+\varphi) - \delta Q_\phi
 \,,\nonumber \\
 Q_{(\gamma)0}
  &=& -Q_{\gamma}(1+\varphi) - \delta Q_\gamma\,,
\eea
where the gravitational redshift (time-dilation) factor $(1+\varphi)$
has been made manifest.

Both the density perturbations, $\delta\rho_\phi$ and $
\delta\rho_\gamma$, and the
curvature perturbation, $\psi$, are in general gauge-dependent.
Specifically they depend upon the chosen time-slicing in an
inhomogeneous universe. The curvature perturbation on fixed time hypersurfaces
is a gauge-dependent quantity: after an arbitrary linear coordinate
transformation, $t\rightarrow t+\delta t$, it transfors as
$\psi\rightarrow \psi+H\delta t$. For a scalar quantity, such as the 
energy density, the corresponding transformation is $\delta\rho\rightarrow
\delta\rho-\dot\rho\delta t$. 
However a gauge-invariant combination can
be constructed which describes the density perturbation on uniform
curvature slices or, equivalently the curvature of uniform density
slices.

The curvature perturbation on uniform total density hypersurfaces,
$\zeta$, is given by \cite{BST}
\begin{equation}
\label{zeta}
\zeta=-\psi-H\frac{\delta\rho}{\dot\rho} \,,
\end{equation}
while the curvature perturbation on uniform inflaton energy density
and radiation energy density
hypersurfaces are respectively  defined as
\begin{eqnarray}
\label{zetaalpha}
\zeta_{\phi}&=&-\psi-H\frac{\delta\rho_\phi}{\dot{\rho_\phi}} \,,
\nonumber\\
\zeta_{\gamma}&=&-\psi-H\frac{\delta\rho_\gamma}{\dot{\rho_\gamma}} \,.
\end{eqnarray}
The total curvature perturbation (\ref{zeta}) is thus a weighted
sum of the individual perturbations
\be
\label{zetatot}
\zeta= 
 \frac{\dot{\rho_\phi}}{\dot\rho}  \zeta_\rho+
\frac{\dot{\rho_\g}}{\dot\rho}  \zeta_\g \,,
\end{equation}
while the difference between the two curvature perturbations
describes a relative gauge-invarariant entropy (or isocurvature) perturbation
 \be
 \label{defS}
  {\cal S}_{\phi\g}=3(\zeta_\phi-\zeta_\g)
   = -3H
\left(
  \frac{\delta\rho_\phi}{\dot{\rho_\phi}}
  - \frac{\delta\rho_\g}{\dot{\rho_\g}} \right) \, .
\end{equation}
From the definitions of the total curvature perturbation
(\ref{zetatot}) and the entropy perturbation (\ref{defS}), we get
for instance that 
\be
 \label{relation}
 \zeta_{\phi}=\zeta+\frac{1}{3} 
  \frac{\dot{\rho_\gamma}}{\dot\rho}{\cal S}_{\phi\gamma}
 \,.
\end{equation}
On wavelengths larger than the horizon scale, 
the perturbed energy conservation equations  for the inflaton energy
density and the radiation energy density can be written, 
including energy transfer, as
\begin{eqnarray} \label{pertenergyexact}
\dot{\delta\rho}_{\phi}+3H(\delta\rho_{\phi}+\delta P_{\phi})
- \left(\rho_{\phi}+P_{\psi}\right)3\dot\psi
&=& Q_{\phi}\varphi+\delta Q_{\phi}\,,\nonumber\\
\dot{\delta\rho}_{\g}+3H(\delta\rho_{\g}+\delta P_{\g})
- \left(\rho_{\g}+P_{\psi}\right)3\dot\psi
&=& Q_{\g}\varphi+\delta Q_{\g}\,.
\end{eqnarray}
The inflaton field and radiation have fixed equations of 
state $(\delta P_\phi=0$ and $\delta P_\g=\delta\rho_\g/3)$ and hence 
there cannot be intrinsic non-adiabatic pressure
perturbations. Using the perturbed $(0-i)$-component
of Einstein's equations  for super-horizon wavelengths
$\dot\psi+H\varphi=-\frac{H}{2}\frac{\delta\rho}{\rho}$, we can 
re-write Eq. (\ref{pertenergyexact})
  in terms of the gauge-invariant curvature
perturbations $\zeta_\phi$ and $\zeta_\g$   \cite{wmu}
\begin{eqnarray}
 \label{dotzetaalpha}
\dot\zeta_\phi &=&
 -{H\left(\delta Q_{\rm{intr},\phi}+
\delta Q_{\rm{rel},\phi}\right)
\over\dot{\rho_\phi}} \,,\nonumber\\
\dot\zeta_\g &=&
 -{H\left(\delta Q_{\rm{intr},\g}+
\delta Q_{\rm{rel},\g}\right)
\over\dot{\rho_\g}} \,,
\end{eqnarray}
where 
\begin{eqnarray}
\label{deltaQintralpha}
\delta Q_{{\rm intr},\phi} &\equiv& \delta Q_\phi -
{\dot{Q}_\phi\over\dot{\rho_\phi}} \delta\rho_\phi \,,\nonumber\\
\delta Q_{{\rm intr},\g} &\equiv& \delta Q_\g -
{\dot{Q}_\g\over\dot{\rho_\g}} \delta\rho_\g\, ,
\end{eqnarray}
are the gauge-invariant instrinsic non-adiabatic energy transfer
perturbations, which are automatically vanishing if
the local energy transfer functions $Q_i$ ($i=\phi,\g$) are 
functions of the local energy density $\rho_i$, and 

\begin{eqnarray}
\label{deltaQrelalpha}
\delta Q_{{\rm rel},\phi} &=&
{Q_\phi\dot\rho \over 2\rho} \left(
{\delta\rho_\phi\over\dot{\rho_\phi}} - {\delta\rho\over\dot\rho}
\right)
= - {Q_\phi \over 6H\rho}  \dot{\rho_\gamma} {\cal S}_{\phi\g}\,,\nonumber\\
\delta Q_{{\rm rel},\g} &=&
{Q_\g\dot\rho \over 2\rho} \left(
{\delta\rho_\g\over\dot\rho_\g} - {\delta\rho\over\dot\rho}
\right)
= - {Q_\g \over 6H\rho}  \dot\rho_\phi {\cal S}_{\g\phi}
\end{eqnarray}
are the gauge-invariant relative non-adiabatic energy transfer 
due to the presence of relative entropy perturbations \cite{wmu}.

\subsection{Perturbing the inflaton decay rate}

Allowing for a perturbed decay rate ($\delta\Gamma\neq 0$) and for a possible
time variation of $\Gamma$ ($\dot\Gamma\neq 0)$, the perturbed
energy transfer is simply given by
\begin{eqnarray}
\delta Q_{\phi}&=&-\Gamma\delta\rho_{\phi}-\delta\Gamma \rho_\phi \,, \\
\delta Q_{\gamma}&=&\Gamma\delta\rho_{\phi}+\delta\Gamma\rho_\phi \,.
\end{eqnarray}
The corresponding  intrinsic non-adiabatic energy
transfer terms from the inflaton are given by

\begin{eqnarray} 
\delta Q_{\rm{intr},\phi}&=& -\delta\Gamma\rho_\phi+\dot\Gamma
\frac{\rho_\phi}{\dot{\rho_\phi}}\delta\rho_\phi\,, \nonumber\\
\delta Q_{\rm{intr},\g}&=& \delta\Gamma\rho_\phi+
\Gamma\delta\rho_\phi-
\Gamma\frac{\dot{\rho_\phi}}{\dot{\rho_\g}}\delta\rho_\phi
-\dot\Gamma
\frac{\rho_\phi}{\dot{\rho_\g}}\delta\rho_\g\,.
\end{eqnarray}
The relative non-adiabatic energy transfer terms are given by
\begin{eqnarray}
\delta Q_{\rm{rel},\phi} &=&
-\frac{\Gamma\rho_{\phi}\dot\rho}{2\rho}\left(
\frac{\delta\rho_{\phi}}{\dot\rho_{\phi}}-\frac{\delta\rho}{\dot\rho}
\right) \,, \\
\delta Q_{\rm{rel},\gamma}&=&
\frac{\Gamma\rho_{\phi}\dot\rho}{2\rho}\left(
\frac{\delta\rho_{\gamma}}{\dot\rho_{\gamma}}-\frac{\delta\rho}{\dot\rho}
\right) \,.
\end{eqnarray}
Thus the evolution equations (\ref{dotzetaalpha}) for the
curvature perturbation on uniform inflaton density hypersurfaces,
$\zeta_\phi$, and uniform radiation density hypersurfaces,
$\zeta_\gamma$, are given by
\begin{eqnarray}
\label{dotzetacurv}
\dot\zeta_\phi
&=&
-{\Gamma\over6} {\rho_\phi\over\rho}
{\dot\rho_\gamma\over\dot\rho_\phi}{\cal S}_{\phi\gamma}+
H\frac{\rho_\phi}{\dot\rho_{\phi}}\delta\Gamma_\phi^{\rm GI} \,, \\
\label{dotzetarad}
\dot\zeta_\gamma
&=&
{\Gamma\over3}
{\dot\rho_\phi\over\dot\rho_\gamma}
\left(1-{\rho_\phi\over2\rho}\right)
{\cal S}_{\phi\gamma} -H\frac{\rho_\phi}{\dot\rho_{\g}}
\delta\Gamma_\g^{\rm GI} \,, \end{eqnarray}
where 

\begin{eqnarray}
\delta\Gamma_\phi^{\rm GI}&=&\delta\Gamma-\dot\Gamma
\frac{\delta\rho_\phi}{\dot\rho_{\phi}}\,,\nonumber\\
\delta\Gamma_\g^{\rm GI}&=&\delta\Gamma-\dot\Gamma
\frac{\delta\rho_\g}{\dot\rho_{\g}}=\delta\Gamma_\phi^{\rm GI}+
\frac{\dot\Gamma}{H}\frac{\dot\rho}{\dot{\rho_\g}}\left(\zeta-\zeta_\phi\right)
\end{eqnarray}
are the gauge-invariant perturbations of the inflaton decay rate. As
anticipated in the Introduction, the gauge-invariant
decay rate perturbation receives contribution from two sources:
the fluctuation of the decay rate $\delta\Gamma$ (generated during the
inflationary stage) and the time variation of the decay rate $\dot\Gamma$
during reheating. 

Writing the total 
curvature perturbation on uniform total density hypersurfaces (\ref{zetatot})
as

\be
\zeta=f\zeta_\phi +\left(1-f\right)\zeta_\g\, ,
\end{equation}
where 
\begin{equation}
f=\frac{3\Omega_\phi+g\Omega_\g}{4\Omega_\g+3\Omega_\phi}\, ,
\end{equation}
we can obtain the following system governing the evolution of the
total curvature perturbation  on uniform total density hypersurfaces 
and the curvature perturbation on uniform inflaton density hypersurfaces

\begin{eqnarray}
\label{master}
\zeta^\prime&=&\left[\frac{\Omega_\phi\left(2g-3\right)}{\left(1-g\right)
\left(4-\Omega_\phi\right)}+\frac{\Gamma^\prime}{H}\frac{(1-g)\Omega_\phi}{4
(1-g)
\left(1-\Omega_\phi\right)-g\Omega_\phi}\right]\left(\zeta-
\zeta_\phi\right)\, ,\nonumber\\
\zeta_\phi^\prime&=&\frac{g\left(4-\Omega_\phi\right)}{2\left(3-2g\right)}
\left(\zeta-
\zeta_\phi\right)-\frac{1-g}{3+g}\frac{\delta\Gamma_\phi^{\rm GI}}{H}\, .
\end{eqnarray} 
In the standard inflationary scenario, $\rho_\gamma$ is supposed to
dominate the initial energy density after reheating and is
assumed to be unperturbed at the end of inflation
(beginning of the reheating stage)  $\zeta_{\gamma,{\rm in}}=0$.
Thus the curvature perturbation is initially (right after the
end of inflation) an adiabatic density perturbation in the inflaton field. 
At this stage some comments are in order.

\begin{itemize}

\item
From the set of equations (\ref{master}), 
it is clear that during reheating, if $
\delta\Gamma_\phi^{\rm GI}=0$, the solution $\zeta=\zeta_\phi$ is a fixed
point attractor.
Therefore, if at the end of inflation 
the  total curvature perturbation  $\zeta$ is 
entirely provided by the curvature perturbation on uniform inflaton 
density hypersurfaces,  
$\zeta_{\rm in}=\zeta_{\phi,{\rm in}}$, the
 total curvature perturbation  $\zeta$ 
remains constant on super-horizon scales. At the end of the reheating 
stage
and beginning of the radiation phase, $\zeta=\zeta_{\gamma}=
\zeta_{\phi,{\rm in}}$.

\item If the inflaton decay rate $\Gamma$ depends only upon the inflaton field,
$\Gamma=\Gamma(\phi)$, then the gauge-invariant perturbation
of the decay rate reads

\begin{equation}
\delta\Gamma_\phi^{\rm GI}=\frac{\partial \Gamma}{\partial\phi}\dot\phi
\left(\frac{\delta\phi}{\dot\phi}-\frac{\delta\rho_\phi}{\dot{\rho_\phi}}
\right)=\frac{\partial \Gamma}{\partial\phi}\dot\phi
\left(\frac{\delta\phi}{\dot\phi}-\frac{\delta\dot\phi}{\ddot\phi}\right)\,.
\end{equation}
Since the long-wavelength solutions for the vacuum
fluctuations in the inflaton field obey the
adiabatic condition $\delta\phi/\dot\phi=\delta\dot\phi/\ddot\phi$, the 
gauge-invariant perturbation
of the decay rate vanishes identically. This implies that 
the source term proportional to $\delta\Gamma_\phi^{\rm GI}$
for $\zeta^\prime_\phi$ 
in Eqs.  (\ref{master}) vanishes identically and 
$\zeta=\zeta_\phi$ is a fixed
point attractor during the reheating stage.
The total curvature perturbation  $\zeta$ 
remains constant on super-horizon scales during reheating.

The same conclusion can be drawn if the decay rate is a function
of the temperature associated to radiation. Suppose that 
the inflaton field is coupled to some   fermion $\psi$
(radiation) 
through the Yukawa coupling ${\cal L}_Y=h\bar\psi\psi\phi$. This
coupling allows the inflaton field to decay into fermions
with a decay rate $\Gamma=\frac{h^2}{8\pi}M_\phi\sqrt{1-
4 M_\psi^2/M_\phi^2}$, 
where $M_\phi$ and $M_\psi$ are the inflaton mass during the
coherent oscillations and the fermion mass, respectively.
During the reheating stage,  at very early times, $t\ll \Gamma^{-1}$, 
the energy density of the universe is dominated by
the scalar field $\phi$ and the radiation 
density is negligible. 
As the scalar field decays into fermions, the decay products rapidly
thermalize forming a plasma with   temperature $T$. The latter
grows  until it reaches a maximum value $T_{MAX}$ and then
decreases as $T\propto a^{-3/8}$ up to the temperature $T_{RH}$ at the
time $t\simeq \Gamma_{\phi}^{-1}$ which determines the end of
reheating \cite{dan,gian}. The thermalized 
fermions, produced during the first stages of reheating, acquire
a plasma mass of the order of $gT$, where $g$ is the typical (gauge) coupling 
governing the fermion interactions \cite{weldon}. This happens because 
forward scatterings of fermions do not change the distribution functions
of particles, but modify their free dispersion relations, producing a 
plasma mass. The decay rate is therefore a function not only of the
inflaton field $\phi$, but also of the temperature
$T\sim \rho_\gamma^{1/4}$. The gauge-invariant decay rate
can be written as

\begin{equation}
\delta\Gamma_\phi^{\rm GI}=\delta\Gamma-\dot\Gamma
\frac{\delta\rho_\phi}{\dot\rho_{\phi}}=\frac{\partial \Gamma}{\partial
\rho_\gamma}\dot{\rho_{\g}}\left(\frac{\dot\rho}{\dot{\rho_\g}}-
\frac{\delta\rho_\phi}{\dot\rho_{\phi}}\right)=\frac{\partial \Gamma}{\partial
\rho_\gamma}\frac{\dot\rho}{H}\left(\zeta-\zeta_\phi\right)\, ,
\end{equation}
which shows that the source term coming from  $\delta\Gamma_\phi^{\rm GI}$
for $\zeta^\prime_\phi$ 
in Eqs.  (\ref{master}) is proportioanl to $\left(\zeta-\zeta_\phi\right)$
and therefore $\zeta=\zeta_\phi$ remains a 
fixed point attractor during the reheating stage. 
The total curvature perturbation remains constant on super-horizon scales.

\item If the inflaton decay rate  depends upon the vacuum expectation value
of another field $\chi$ whose quantum fluctuations are excited during inflation
and whose time variation  during reheating
is negligible, $\dot \chi\simeq 0$, one recovers the inhomogeneous
reheating scenario  \cite{gamma1,gamma3} 
with $\delta\Gamma_\phi^{\rm GI}=\delta\Gamma=\left(
\partial\Gamma/\partial \chi\right)\delta \chi$ and $\dot\Gamma=0$. 
During the reheating stage, when the inflaton field is still oscillating
around the minimum of its potential and $t\ll \Gamma^{-1}$, 
we can set $\Omega_\phi\simeq 1$, $g\simeq 0$. 
Eqs. 
(\ref{master}) reduce to
\begin{eqnarray}
\label{mastera}
\zeta^\prime&\simeq &-\left(\zeta-
\zeta_\phi\right)\, ,\nonumber\\
\zeta_\phi^\prime&\simeq&-\frac{1}{3}\frac{\delta\Gamma}{H}\, ,
\end{eqnarray} 
which is easily solved to give (going back to cosmic time)

\begin{equation}
\zeta\simeq \zeta_{\rm in}-\frac{1}{3}\int_{t_{\rm in}}^{t}\, dt^\prime
\frac{H\left(t^\prime\right)}{a\left(t^\prime\right)}\,
\int_{t_{\rm i}}^{t^\prime}\,dt^{\prime\prime}\,H\,\left(t^{\prime\prime}\right)
a\left(t^{\prime\prime}\right)\,\frac{\delta\Gamma}
{H\left(t^{\prime\prime}\right)}\,,
\end{equation}
where $t_{\rm in}$ denotes the initial time  of the reheating stage.
Since on super-horizon scales we may consider $\delta\Gamma$ as a constant
and $a\propto t^{2/3}$ during the inflaton coherent oscillation phase, 
we obtain at the end of the reheating phase (which we set to be
at $t= \Gamma^{-1}$)

\begin{equation}
\zeta\simeq \zeta_{\rm in}-\frac{2}{15}\frac{\delta\Gamma}{\Gamma}\, .
\end{equation}
Since deep in the radiation phase the gravitational potential is
$\psi_\gamma=-\frac{2}{3}\zeta$, we obtain 

\begin{equation}
\psi_\gamma=\frac{4}{45}\frac{\delta\Gamma}{\Gamma} 
\simeq \frac{1}{9}\frac{\delta\Gamma}{\Gamma}\, ,
\end{equation}
which confirms the findings of Refs. \cite{gamma1,gamma3}, in the case 
where the initial curvature perturbation $\zeta_{\rm in}$ is tiny.

\end{itemize}

\subsection{Effects of a time variation of the inflaton decay rate}

Let us now analyze the new source of the variation of the total curvature
perturbation on large scales proportional to $\dot\Gamma$.
We suppose that
the decay rate is a function of time during the reheating stage, $\dot\Gamma
\neq 0$, 
and that its fluctuations during inflation are vanishing, $\delta\Gamma=0$.
Under these circumstances, the gauge-invariant decay rate perturbation
is given by 

\begin{equation}
\delta\Gamma_\phi^{\rm GI}=-\dot\Gamma
\frac{\delta\rho_\phi}{\dot\rho_{\phi}}\,.
\end{equation}
Solving Eqs. (\ref{master}) (going back to
cosmic time), we find the total curvature perturbation at the end of the
reheating stage is given by 

\begin{equation}
\label{n}
\zeta_{\rm f}=\zeta_{\rm in}-\int_{t_{\rm in}}^{t}\,dt^\prime\,
f(t^\prime)\,e^{\int_{t_i}^{t}(f-g) dt^\prime}\,\int_{t_{\rm in}}^{t^\prime}
\,dt^{\prime\prime}
\, S(t^{\prime\prime})\, e^{-\int_{t_i}^{t^\prime}(f-g) dt^{\prime\prime}}\, ,
\end{equation}
where

\begin{eqnarray}
f&=&H\left[\frac{\Omega_\phi\left(2g-3\right)}{\left(1-g\right)
\left(4-\Omega_\phi\right)}+\frac{\dot\Gamma}{H^2}\frac{(1-g)\Omega_\phi}{4
(1-g)\left(1-\Omega_\phi\right)-g\Omega_\phi}\right]\, ,\nonumber\\
g&=&\frac{g\left(4-\Omega_\phi\right)}{2\left(3-2g\right)}\,H \, ,\nonumber\\
S&=&\dot\Gamma\,\frac{1-g}{3+g}\, 
\frac{\delta\rho_\phi}{\dot\rho_{\phi}}\,.
\end{eqnarray}
Let us give an example. 
Suppose again that the inflaton field decays into some  fermion $\psi$
(radiation) 
through the Yukawa coupling ${\cal L}_Y=h\bar\psi\psi\phi$. The fermion
mass $M_\psi$ is a function of a scalar field $S$, 

\begin{equation}
M_\psi=\lambda S\, .
\end{equation}
The potential of the scalar field $S$ reads $V(S)=\frac{1}{2}M_S^2 S^2+
C^2\frac{H^2}{2}\left(S-S_0\right)^2$, with $C\gg 1$. 
If during inflation $H\gg M_S$, the
field $S$ is sitting at the position $S=S_0$ and its quantum fluctuations
are not excited since its effective mass $\sim CH$ is much larger than
the Hubble rate. At the end of inflation, the inflaton starts oscillating
with mass $M_\phi$.  
If $\lambda S > M_\phi/2$, the inflaton 
cannot decay since  the mass of the fermion $M_\psi$ is larger than
$M_\phi/2$ and the decay
channel into fermions is kinematically
forbidden. However, during the coherent 
oscillation phase, 
the Hubble
parameter drops down as $a^{-3}$. If the Hubble rate becomes  smaller
than $C^{-1} M_S$,  the scalar field $S$ may start rolling
down towards the minimum at $S=0$ and settle there, 
thus decreasing the 
fermion mass.
As soon as $M_\psi=\lambda S$ becomes smaller than $M_\phi/2$, say at some 
time
$t=t_*$,  the inflaton decay channel into
fermions becomes accessible and the decay  rate rises from zero
to a nonvanishing value $\Gamma_0=\frac{h^2}{8\pi}M_\phi$.  We can 
safely approximate the 
decay rate as $\Gamma(t)=\Gamma_0\,\theta\left(t-t_*\right)$
corresponding to 

\begin{equation}
\dot\Gamma=\Gamma_0\,\delta\left(t-t_*\right)\,.
\end{equation}
From Eqs. (\ref{master}), we expect that the total curvature
perturbation suffers a jump at $t=t_*$.  
Inserting this expression in Eq. (\ref{n}) and working in the
limit $\Gamma_0/H_*\la 1$,  we find that the total curvature perturbation
jumps from the initial value $\zeta_{\rm in}$ to the final value

\begin{equation}
\zeta_{\rm f}
\simeq \left(1+\frac{4}{45}\,\frac{\Gamma_0}{H_*}\right)\,\zeta_{\rm in}
\, .
\end{equation}
In the opposite limit $\Gamma_0/H_*\ga 1$,  we find

\begin{equation}
\zeta_{\rm f}
\simeq \left(1+\frac{4}{5}\,\frac{H_*}{\Gamma_0}\right)\,\zeta_{\rm in}
\, .
\end{equation}
One can envisage the alternative possibility that during the reheating stage
the Hubble rate remains always larger than $C^{-1} M_S$. Under these
circumstances, the field $S$ does not roll towards $S=0$, but remains stuck
to the minimum of its potential at 
$S=\frac{CH^2}{M_S^2+CH^2}S_0$.
The location of the minimum, however, changes adiabatically with time,
$\dot S/S\sim -6 (M_S^2/C^2 H^2)$. Correspondingly, the decay rate
$\Gamma=\Gamma_0\sqrt{1-\frac{4 M_\psi^2}{M_\phi^2}}$ changes with time
as $\dot\Gamma/\Gamma_0\sim 48\frac{\lambda^2 S_0^2 M_S^2}{C^2 M_\phi^2 H}$.
Solving Eq. (\ref{n}) gives 

\begin{equation}
\zeta_{\rm f}\simeq \left[1-\frac{2}{15}\,\left(\frac{\dot\Gamma}{\Gamma_0^2}
\right)_{\Gamma_0=H}
\right]\,\zeta_{\rm in}
\simeq \left[1-\frac{96}{15}\,\frac{\lambda^2 S_0^2 M_S^2}{C^2 M_\phi^2 
\Gamma_0^2}
\right]\,\zeta_{\rm in}
\, .
\end{equation}
We conclude that during the reheating stage the total curvature perturbation
may be altered on super-horizon scales if the decay rate of the
inflaton changes with time. Even though the amount of change is model
dependent, the shift in the total curvature perturbation is 
always proportional to the value of the total curvature perturbation at the
end of inflation and the beginning of the reheating stage $\zeta_{\rm in}$.

\subsubsection{Violation of the  
consistency relation \label{consistencysection}}

During inflation both scalar and tensor perturbations are generated.
As we have seen in the previous section, the total curvature
perturbation can change on super-horizon scale during reheating if the
inflaton field decay rate 
is a function of time. The difference between the values
of the total curvature
perturbation at the end of reheating and at the beginning of inflation
can be parametrized by  $\Delta \equiv \left(\zeta_{\rm f}-\zeta_{\rm in}
\right)/\zeta_{\rm in}$.
The power
spectrum of scalar curvature perturbations at the end of the 
reheating stage and the beginning of the radiation phase
is therefore given by

\begin{equation}
{\cal P}_\zeta (k) = \frac{k^3}{2 \pi^2} \left|\zeta_{\rm f} \right|^2 =
\frac{k^3}{2 \pi^2} \left(1+\Delta\right)^2\left|\zeta_{\rm in} \right|^2= 
\frac{1}{8\pi^2}\frac{\left(1+\Delta\right)^2}
{\epsilon}\left(\frac{H(k)}{M_{P}}\right)^2 \, ,
\label{cur}
\end{equation}
where $M_{P}$ is the Planck mass, $H(k)$ indicates the value of the Hubble 
parameter when a given
wavelength $\lambda=2\pi/k$ crosses the horizon, {\it i.e.}, when $k=aH$, 
and $\epsilon = -\dot{H}/{H^2}$ 
is a slow-roll parameter accounting 
for the time variation of the Hubble rate during inflation.

The primordial spectrum of
gravitational waves is given by 

\begin{equation}
{\cal P}_T(k) =
\frac{2}{\pi^2}\left(\frac{H(k)}{M_{P}}\right)^2\, .
\label{psh}
\end{equation}
From expressions (\ref{cur}) and (\ref{psh}), we can predict
a consistency relation 
between the amplitude of the scalar perturbations, ${\cal
P}_\zeta(k)$, the amplitude of the tensor perturbations, 
${\cal P}_T(k)$, and the
tensor spectral index, $n_T\equiv d\ln {\cal P}_T(k)/d\ln k$.
Indeed, since ${\cal P}_T(k)\propto H^2(k)$, $n_T$ is given
by $n_T = d\ln H^2(k)/d\ln k = -2\epsilon$ and 
the consistency relation, for $\vert \Delta \vert \ll 1$, reads
\begin{equation}
 \frac{{\cal P}_T(k)}{{\cal P}_\zeta(k)}= 16\left(1-2\Delta\right) 
\epsilon = -8\left(1-2\Delta\right) n_T\, .
\end{equation}
This consistency relation differs from the traditional one,
${\cal P}_T/{\cal P}_\zeta= -8n_T$, obtained for one-single field models of
inflation \cite{reviewrocky} when the inflaton decay rate during reheating
does not change with time. Notice that, if $\Delta<0$, the 
tensor-to-scalar amplitude ratio can be larger than 
for one-single field models of
inflation.  
Departures from the traditional consistency relation can be caused by
other reasons: higher-order terms in the expansion in slow-roll 
parameters \cite{reviewrocky}, quantum loop corrections \cite{kl}
or the presence of  multiple fields during inflation \cite{noi}. Given
CMB B-mode polarization measurements, departures from the 
traditional consistency relation can be detected
if the tensor-to-scalar amplitude ratio is larger than about
$10^{-3}$ \cite{kn} and would rule
out the simplest case of one-single field models of inflation with
no variation of the decay rate.

\section{The nonperturbative inflaton decay} \label{sec:pre}

As we already mentioned, if the classical inflaton field $\phi$ 
very rapidly (explosively) decays into either its own quanta  
or into other bosons due to broad parametric resonance, the so-called
preheating stage \cite{pre}, Eqs. (\ref{defbackQb}) should be modified.

Suppose the inflaton is coupled to a light scalar $\chi$ with coupling
${\cal L}=\frac{1}{2}g^2\phi^2\chi^2$. During the coherent oscillations of
the inflaton field, the $\chi$-quanta satisfy the so-called Mathieu
equation which leads to the existence of exponential instabilities
$\chi_k\propto e^{\mu_k^{(n)}M_\phi t}$ within the set of resonance
bands of frequencies $\Delta k^{(n)}$ labelled by an integer $n$
and (for the first few bands)

\begin{eqnarray}
\mu^{(n)}&=&\frac{M_\phi}{2n}\frac{q^n}{\left(2^{n-1}(n-1)!\right)^2}\, ,
\nonumber\\
q&=&\frac{g^2\phi^2}{M_\phi^2}\, .
\label{mu}
\end{eqnarray}
For $q\ltsim 1$,  preheating occurs mainly
in the first  resonance
band with $\mu^{(1)}=qM_\phi/2$ and a width $\Delta k^{(n)}\sim\mu^{(1)}$. 
For $q\gg 1$, however, many resonance bands are excited and preheating
occurs in the broad resonant regime, with $\mu_k^{(n)}\sim 0.17$, 
independent of $q$. At the beginning of
the preheating stage, the so-called linear stage, 
the $\chi$-quanta grow exponentially in time till
back-reaction effects set in. They are originated by the fact that
the inflaton effective mass squared 
$M^2_{\rm eff}=M_\phi^2+g^2\langle \chi^2\rangle$
becomes dominated by the second term, thus decreasing the parameter $q$, and
by the scattering of the produced particles off the zero mode 
\cite{kt1,kt2,tow}.

During the linear stage, it is a good approximation to
write the continuity equations (\ref{m}) as

\begin{eqnarray}
\dot\rho_{\phi}
&=&-3H\left(\rho_{\phi}+P_{\phi}\right) -\mu\rho_\gamma\,,\nonumber\\
\dot\rho_{\gamma}
&=&-3H\left(\rho_{\gamma}+P_{\gamma}\right) +\mu\rho_\gamma\,,
\label{mp}
\end{eqnarray}
where we have indicated by $\rho_\gamma$ the energy density
of light degrees of freedom, the $\chi$-particles generated during the
first stage of preheating and by $\mu$ the rate of production. Notice that
$\mu$ is a function of time as clear from Eqs. (\ref{mu}). 

The background equations (\ref{m}) can be re-written as

\begin{eqnarray}
\label{omegasprimep}
 \Omega_\phi' &=& \Omega_\gamma
 \left(\Omega_\phi-r\right) \label{dinp3} \,,\\
 \Omega_\gamma' &=& r\Omega_\gamma\Omega_\phi \label{dinp1}\,,\\
\label{gprimep}
 r' &=& r\frac{\mu^\prime}{\mu}+{r\over2}(4-\Omega_\phi) \label{dinp4}\,,
\end{eqnarray}
where we have set

\begin{equation}
r=\frac{\mu}{H}\, .
\end{equation}
The local energy transfer functions are now given by $Q_\phi=-\mu \rho_\gamma$
and $Q_\gamma=\mu\rho_\gamma$. Proceeding as in 
subsection \ref{sec:gauge}, we
can write  the equations for the change of the 
curvature perturbation on uniform inflaton energy density
and radiation energy density on super-horizon scales

\begin{eqnarray}
\label{dotzetacurvp}
\dot\zeta_\phi
&=&
{\mu\over3} {\dot{\rho}_\gamma\over\dot{\rho}_\phi}
\left(1-\frac{1}{2}\frac{\rho_\gamma}{\rho}\right){\cal S}_{\phi\gamma}+
H\frac{\rho_\gamma}{\dot\rho_{\phi}}\delta\mu_\phi^{\rm GI} \,, \\
\label{dotzetaradp}
\dot\zeta_\gamma
&=-&
{\mu\over6}\frac{\rho_\gamma}{\rho}
{\dot\rho_\phi\over\dot\rho_\gamma}
{\cal S}_{\phi\gamma} -H\frac{\rho_\gamma}{\dot\rho_{\g}}
\delta\mu_\g^{\rm GI} \,, \end{eqnarray}
where 

\begin{eqnarray}
\delta\mu_\phi^{\rm GI}&=&\delta\mu-\dot\mu
\frac{\delta\rho_\phi}{\dot\rho_{\phi}}\,,\nonumber\\
\delta\mu_\g^{\rm GI}&=&\delta\mu-\dot\mu
\frac{\delta\rho_\g}{\dot\rho_{\g}}=\delta\mu_\phi^{\rm GI}+
\frac{\dot\mu}{H}\frac{\dot\rho}{\dot{\rho_\g}}\left(\zeta-\zeta_\phi\right)
\end{eqnarray}
are the gauge-invariant perturbations of the inflaton decay rate during
the linear stage of preheating. 

The equations for the total curvature perturbation and the 
curvature perturbation on uniform inflaton energy density become

\begin{eqnarray}
\label{masterp}
\zeta^\prime&=&\left\{\Omega_\phi\left(2r-\frac{3}{2}\right)
-\frac{3(\Omega_\phi-r)(r-4)+\frac{1}{2}r(4-\Omega_\phi)-
\frac{\mu^\prime}{\mu}r\left[\Omega_\gamma(\Omega_\phi-4)-1\right]
}{(r-4)(4-\Omega_\phi)}\right\}
\left(\zeta-
\zeta_\phi\right)\, ,\nonumber\\
\zeta_\phi^\prime&=&-\frac{r\left(4-\Omega_\phi\right)}{(3-r)\Omega_\phi+r}
\left(1-\frac{1}{2}\Omega_\gamma\right)
\left(\zeta-
\zeta_\phi\right)-\frac{\Omega_\gamma}{(3-r)\Omega_\phi+r}
\frac{\delta\mu_\phi^{\rm GI}}{H}\, .
\end{eqnarray} 
From Eqs. (\ref{masterp}), we conclude that 
if the resonance
parameter $\mu$ has either spatial or time variations, then the total
curvature perturbation can be modified on large-scales. As a 
matter of fact, the parameter $\mu$ is far from being 
constant during preheating. This fact was not appreciated
in previous studies  of the effects of preheating
on super-horizon perturbations \cite{bf}.
If the nonperturbative 
inflaton decay rate $\mu$ depends only upon the inflaton field,
$\mu=\mu(\phi)$, 
then the gauge-invariant perturbation
of the decay rate $\delta\mu_\phi^{\rm GI}$ vanishes identically and 
$\zeta=\zeta_\phi$ is a fixed
point attractor during the preheating stage.
The total curvature perturbation  $\zeta$ 
remains constant on super-horizon scales during preheating.
This happens, for instance, if Eq. (\ref{mu}) holds and $\mu\sim q^n\sim\
\phi^{2n}$. However, one can envisage other possibilities. 
For instance, as in the inhomogeneous reheating scenario, the
coupling constant $g^2$ between the inflaton field $\phi$ and the
light bosons $\chi$ can be a function of some other field which remains
light during inflation. This leads to spatial fluctuations of the
$q$-parameter. Tiny variations of $q$ lead
to large variations of the efficiency in extracting energy out of the 
inflaton zero mode \cite{kt2} and may have a dramatic effect
on the final temperature fluctuations after the system thermalizes.
This analysis will be presented in a separate publication 
\cite{inprep}.

\section{Conclusions}  \label{sec:conc}

We have studied the evolution of large-scale curvature
perturbations during the reheating stage after inflation in the case
in which spatial and temporal variations of the inflaton decay rate are
present. We have shown that 
the  total curvature perturbation
$\zeta$  can change on large scales due to either spatial variations of
the decay rate originated during inflation -- the so-called inhomogeneous
reheating scenario \cite{gamma1,gamma3} --
or to a time variation of the decay rate during reheating. 
If only the latter source is present, 
the final curvature perturbation at the beginning of the radiation phase
can be either smaller or larger than the curvature perturbation
at the beginning of the reheating stage. This result leads to a violation
of the consistency relation for one-single field models of inflation
and to the observationally 
promising possibility that the tensor-to-scalar 
amplitude ratio
is larger than in the standard scenario.
Finally, we have presented the equations for the evolution
of the total curvature perturbation and the 
curvature perturbation on uniform 
inflaton energy density
hypersurfaces in the case in which the vacuum energy driving inflation is
released into radiation through a (linear) stage of preheating.

\vskip 0.5cm

\centerline{\bf Acknowledgements}
\vskip 0.2cm
It is  a pleasure to thank R. Kolb and I. Tkachev for discussions.


\end{document}